\documentclass[aps,prb,superscriptaddress,floatfix, twocolumn,showpacs]{revtex4}
\usepackage{amsmath}
\usepackage{amsfonts}
\usepackage{amssymb}
\usepackage{color}
\usepackage{ dsfont }
\newcommand{\sgn}{\text{sgn}}
\newcommand{\beq}{\begin{equation}}
\newcommand{\eeq}{\end{equation}}
\newcommand{\beqa}{\begin{eqnarray}}
\newcommand{\eeqa}{\end{eqnarray}}
\usepackage{graphicx,amssymb,amsmath,subfigure}
\usepackage{dcolumn}
\usepackage{bm}

\newcommand{\para}[1]{\left(#1\right)}
\newcommand{\abs}[1]{\left|#1\right|}

\begin{document}
\title{Topological electric current from time--dependent elastic deformations in graphene}
\author{Abolhassan Vaezi}
\affiliation{Department of Physics, Cornell University, Ithaca, New York 14853, USA}
\affiliation{School of Physics, Institute for Research in Fundamental Sciences, IPM, Tehran, 19395-5531, Iran}
\author{Nima Abedpour}
\affiliation{School of Physics, Institute for Research in Fundamental Sciences, IPM, Tehran, 19395-5531, Iran}

\author{ Reza Asgari}
\email{asgari@ipm.ir}
\affiliation{School of Physics, Institute for Research in Fundamental Sciences, IPM, Tehran, 19395-5531, Iran}
\author{Alberto Cortijo}
\affiliation{Instituto de Ciencia de Materiales de Madrid,
CSIC, Cantoblanco; 28049 Madrid, Spain.}
\author{Mar\'ia A. H. Vozmediano}
\affiliation{Instituto de Ciencia de Materiales de Madrid,
CSIC, Cantoblanco; 28049 Madrid, Spain.}
\date{\today}

\begin{abstract}
We show the possibility of inducing an edge charge current by
applying time--dependent strain in gapped graphene samples
preserving time reversal symmetry. We demonstrate that this edge
current has the same origin as the valley Hall response known to
exist in the system.
\end{abstract}

\maketitle

\section{ Introduction.}
%

Topological insulators~\cite{HK10,QZ11} are a hallmark of the
condensed matter physics of the 21st century. They realize a new
state of matter characterized by topology rather than symmetry.
The topologically non trivial character is strongly related to the
discrete symmetries of the band Hamiltonian like time reversal
symmetry (TRS), inversion, or parity. It has observable
consequences in the form of non dissipative currents at the edge
of the  sample  or, equivalently, quantized transverse electric
responses to external electromagnetic probes like the quantum Hall
conductivity, $\sigma_{xy}$.~\cite{KDP80}

The quantum Hall example led to the assumption that breaking TRS
was an essential ingredient for the observation of topological
phenomena which will not occur in insulators  preserving TRS. The
reason is clear: the Hall conductivity is proportional to the
integral over the Brillouin zone of the Berry connection which is
zero in time reversal invariant systems \cite{HK10}. The proposals
of Semenoff and  Haldane  \cite{S84,H88} of getting Landau levels
in a system with zero applied magnetic field followed by the
description of the quantum spin Hall effect  \cite{KM05,BZ06}
paved the way to the development of the actual field of
topological insulators. The spin Hall effect was the first example
of a topological  response in a TRS invariant system and is based
on the recognition that the presence of additional degrees of
freedom (spin in this case) allows to define different types of
masses  that give rise to other topological currents.~\cite
{RMetal09,CGV10}


Graphene~\cite{Netal05,Zetal05,CGetal09}, the best example of
Dirac fermions with extra quantum numbers (spin, valley, layer) is
the ideal model to test this type of quantized responses. The
neutral system  has two  inequivalent Fermi points (valleys)
located at the corners of the Brillouin zone. Because of their
large separation in momentum space, inter--valley scattering is
strongly suppressed and in the absence of short range disorder or
interactions the valley index remains a good quantum number. In
these circumstances a valley Hall effect can occur similar to the
spin Hall effect where carriers in different valleys flow to
opposite transverse edges driven by an in-plane external electric
field. The valley Hall effect was already discussed in the early
times of graphene~\cite{XYN07} and ``Valleytronics" applications
were proposed.~\cite{RTB07} Valley currents induced by AC fields
or optical radiation have been experimentally realized in various
materials.~\cite{KTetal11,ZDetal12,MHetal12}

One of the most interesting aspects of graphene is the tight
relation between electronic excitations and mechanical
deformations of the lattice. In the very successful tight
binding--elasticity approach, lattice deformations couple to the
electronic current in the form of gauge fields and scalar
potentials very similar to the usual electromagnetic gauge
potential.~\cite{SA02,VKG10} Time dependent strains give rise to a
``synthetic" electric field  that will play a major role in the
present work. In the valley Hall effect there is no charge
accumulation at the edges because the external electric field
couples with the same sign to both valleys. We will show that a
charge current can be generated from time-dependent elastic
deformations not breaking TRS in graphene. The result lies on a
mixed Chern Simons term in the effective action  that involves the
electromagnetic and the elastic vector potentials. The interplay
of strain and valley physics have been explored
previously~\cite{GKV08,GKG10,ZB11} and some consequences of having
time--dependent strain in graphene were
considered.~\cite{OGM09,FF12,JLetal13,ICetal13}

The paper is organized as follows. In Sec. II we introduce the
formalism that will be used in calculating topological current. In
Sec. III we present our analytical and numerical results for
suggesting an experimental realization in time-dependent strained
graphene sheets. Sec. IV contains discussions and conclusions.

\section{The model and Theory}

In the absence of lattice deformations the low energy electronic
degrees of freedom around the two Fermi points in graphene can be
described  by a massless Dirac Hamiltonian \cite{CGetal09}:
\begin{equation}
H(\mathbf{k})=\sum_{\tau}\psi^{+}_{\tau,\mathbf{k}}\left(\tau\sigma_{x}k_{x}+\sigma_{y}k_{y}\right)\psi_{\tau,\mathbf{k}},\label{masslessham}
\end{equation}
where $\tau=\pm1$ refers to the two Fermi points and
$\psi_{\tau,\mathbf{k}}$ represents two species of spinors. The
Fermi velocity will not play a role in our discussion and has been
put to one as well as $\hbar$. The time reversal symmetry
operation interchanges the two species of spinors, keeping
Eq.~(\ref{masslessham}) time reversal invariant. An essential
ingredient for the quantized response is the presence of a gap in
the spectrum. In the quantum Hall effect the insulating behavior
is induced by the perpendicular magnetic field. When spins are
neglected, there are essentially four ways of opening a gap in the
otherwise linear spectrum of Eq.~\eqref{masslessham}.~
\cite{CGV12} Three of them are time reversal invariant and
physically correspond to inducing a different on site potential to
the two sublattices  or coupling the degrees of freedom to a
Kekul\'{e} distortion. The corresponding mass has the same sign
for both Fermi points. The fourth one breaks time reversal
symmetry and was used by Haldane in his proposal for the anomalous
quantum Hall effect in the honeycomb lattice.~\cite{H88} We will
restrict ourselves to the first mass mentioned above, \beq
H_{m}=m\psi^{+}_{\tau,\mathbf{k}}\sigma_{3}\psi_{\tau,\mathbf{k}},
\label{mass} \eeq and discuss later on the possible mechanism to
generate this term in real samples.

As discussed extensively in Ref.~[\onlinecite{VKG10}], a
deformation of the graphene lattice gives rise to a fictitious
gauge field \beq {\bf A}^{el}= {\kappa \Phi_0 \over \pi
}\begin{pmatrix}
u_{xx}-u_{yy} \\
-2u_{xy}
\end{pmatrix}
\label{gaugefield} \eeq where $\kappa \simeq 3$ nm$^{-1}$,
$\Phi_0$ is the flux quantum and $u_{ij}$ is the strain tensor
which can be written as $u_{ij} = \frac{1}{2}[\partial_j
u_i+\partial_i u_j+(\partial_i h)(\partial_j h)]$ in terms of the
in-plane and out-of-plane displacements $\mathbf u$ and $h$
respectively. This field couples minimally to the electronic
excitations with opposite signs to the two valleys. Hence in the
presence of an external electromagnetic and elastic field the
interacting Hamiltonian reads:
\begin{eqnarray}\label{H-a}
H_{A}=&-&\sum_{\tau}[\psi^{+}_{\tau,\mathbf{k}}\tau\sigma_{x}\left(eA^{em}_{x}+\tau\hat{\beta} A^{el}_{x}\right)\psi_{\tau,\mathbf{k}}]-\nonumber\\
&-&\sum_{\tau}[\psi^{+}_{\tau,\mathbf{k}}\sigma_{y}\left(eA^{em}_{y}+\tau\hat{\beta} A^{el}_{y}\right)\psi_{\tau,\mathbf{k}}],
\label{intham}
\end{eqnarray}
where $e$ is the electric charge, $\mathbf{A}^{em}$ and
$\mathbf{A}^{el}$ stand for the electromagnetic and the elastic
vector fields respectively, and we have encoded the strength of
the elastic coupling  in the parameter $\hat{\beta}>0$. Note that
we multiply $\mathbf{A}^{el}$ by $\tau$ in Eq.~\eqref{H-a} since
the two valleys couple with opposite charges to strain. This is
due to the fact that the strain gauge field, $\mathbf{A}^{el}$,
respects the time reversal symmetry under which the $K$ and the
$K'$ valleys will be interchanged.

\subsection{The topological current}

In order to obtain the topological response of a gapped system to
external gauge fields, we must find its topological indices. For
example, if quasiparticles of the system couple to a gauge field
with charge $q$, the off-diagonal conductance of the system
depends on the first Chern number, $C$, as
$\sigma_{xy}=C{q^2}/{h}$. The Chern number can itself be computed
through integrating the Berry curvature of the ground-sate over
the momentum space. For example, the Burry curvature for a
two-band system with
$\mathcal{M}_{k}=\vec{d}_{k}\cdot\vec{\sigma}$ Hamiltonian
reads~\cite{Wen04}
\begin{eqnarray}\label{Berry}
\mathcal{F}_{xy}=\frac{1}{2}\hat{d}_{k}.\para{\partial_{k_x}\hat{d}\times \partial_{k_y}\hat{d}}.
\end{eqnarray}
where $\hat{k}={\vec{k}}/{\abs{k}}$. Since, a gapped graphene can
be viewed as the two massive Dirac cones, we first compute the
response of a single one. Consider a generic Dirac cone with
$\mathcal{M}_{k}=v_{\rm
F}\para{\eta_{x}k_{x}\sigma_x+\eta_{y}k_{y}\sigma_y}+m\sigma_{z}$
Hamiltonian, where $\eta_x$, and $\eta_y$ take $\pm 1$ values.
Therefore, $\vec{k}=\para{\eta_{x}v_{\rm F}k_x,\eta_{y}v_{\rm
F}k_y,m}$. Using the corresponding  $\hat{k}$ and plugging it in
Eq.~\eqref{Berry}, the Chern number reads
\begin{eqnarray}\label{Chern-1}
C=\frac{1}{2}\sgn\para{m\eta_x \eta_y}.
\end{eqnarray}

Therefore, in the gaped system with the mass term given in
Eq.~\eqref{mass} the band structure around each Fermi point is
topologically characterized by a Chern number which takes opposite
values at the two Fermi points (due to the time reversal symmetry)
~\cite{Wen91,Wen04}: \beq
C_{\mathbf{K}}=-C_{\mathbf{K}'}=\sgn(m)/2. \eeq In the absence of
elastic deformations as a response to an external electric field
${\bf E}^{em}$ the induced charge current at each Fermi point is
\beq \langle
J^{i}_{\tau}\rangle=e^{2}C_{\tau}\varepsilon^{ij}E^{em}_{j}, \eeq
so the total charge current $\langle
J^{i}_{\mathbf{K}}+J^{i}_{\mathbf{K}'}\rangle$ vanishes. However
there is still a topological response encoded in the quantity
$\langle J^{i}_{\mathbf{K}}-J^{i}_{\mathbf{K}'}\rangle$ which is
not zero and physically represents a current imbalance between the
two Fermi points, this is the manifestation of the quantum valley
Hall effect.

Consider now a time--dependent elastic deformation described by a vector field $\mathbf{A}^{el}(t)$ as the one
described in the previous section.
Its associated synthetic electric field
$E^{el}_{j}=\partial_t A^{el}_j(t)$ will couple with opposite signs to the two valleys. Hence
this type of deformation 
will induce a charge response in the system at each Fermi point:
\beq
\langle J^{i}_{\tau}\rangle\sim\tau C_{\tau}\varepsilon^{ij}E^{el}_{j}.
\eeq
 Now, because $C_{\mathbf{K}}=-C_{\mathbf{K}'}$ the total net charge current is  non zero and its value is twice larger than the value at each Fermi point:
\beq
\langle J^{i}_{\mathbf{K}}+J^{i}_{\mathbf{K}}\rangle\sim2C_{\mathbf{K}}\varepsilon^{ij}E^{el}_{j}.
\eeq
\begin{figure}[t]
\begin{center}
\includegraphics[angle=0,width=0.8\linewidth]{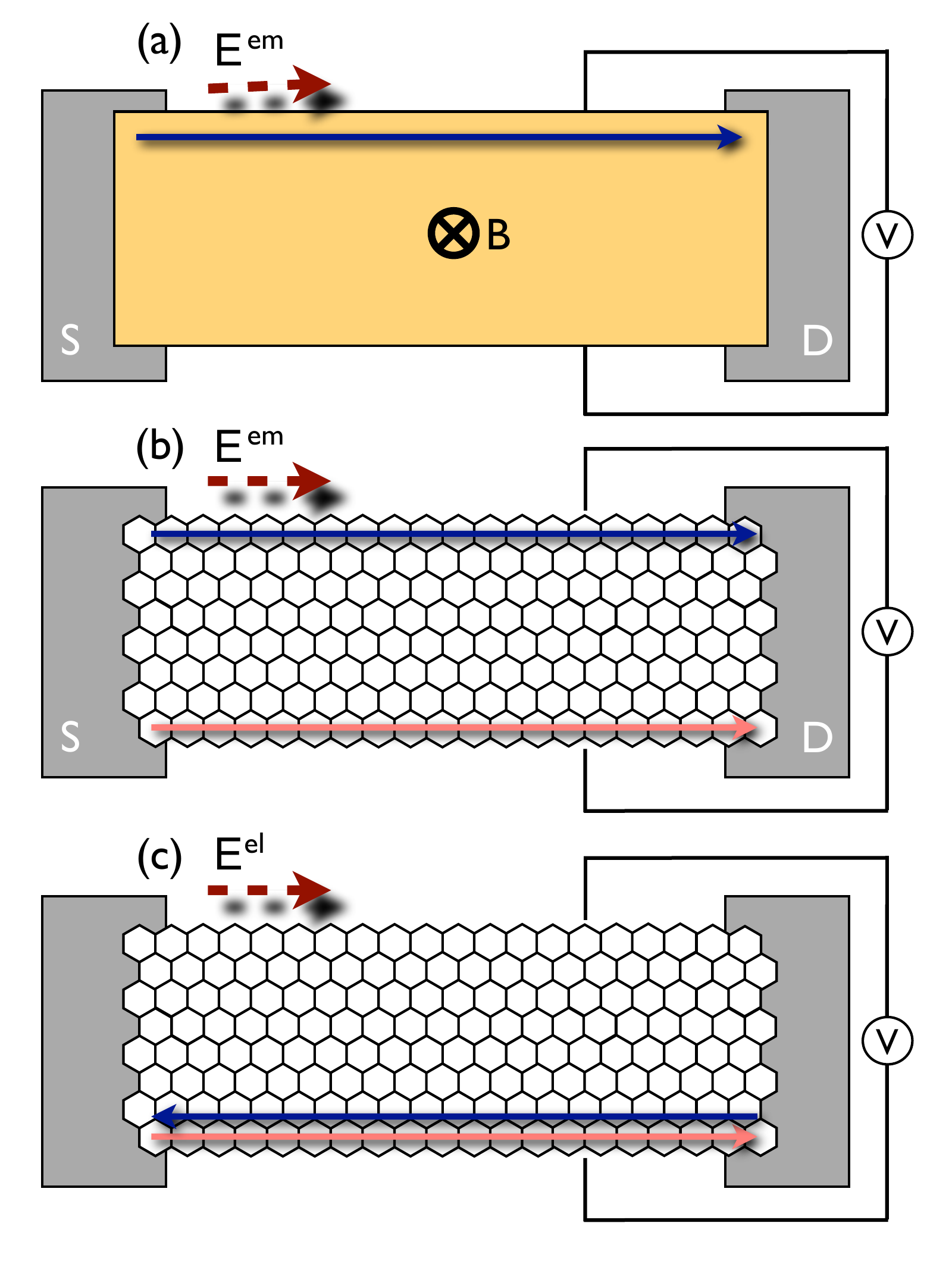}
\end{center}
\caption{(color online) Comparison between the Hall effect (upper), the valley Hall effect (middle), and the effect proposed in the text (lower) following the scheme of the original implementation of the experiment first proposed by Hall. The arrows indicate the flow of valley polarized electrons under the action of an external electric field or voltage (upper and middle) and of a time dependent strain (lower part). In the later case the total current is
$<J^i_K+J^i_{K'}>\sim E^{el}_j$.}
\label{fig1}
\end{figure}
We can make this statement more formal by considering the
effective background field theory.~\cite{CGV10} Within a
functional integral approach one can integrate out the fermionic
degrees of freedom $\psi_{\tau,\mathbf{k}},
\psi^{+}_{\tau,\mathbf{k}}$ in the action derived from Eq.
\eqref{intham} and write the odd part of the effective Lagrangian:
\begin{eqnarray}
\mathcal{L}_\mathit{eff}&=&2 e\hat{\beta} C_{\mathbf{K}}\varepsilon^{\mu\rho\nu}A^\mathit{em}_{\mu}\partial_{\rho}A^\mathit{el}_{\nu}+2e\hat{\beta} C_{\mathbf{K}}\varepsilon^{\mu\rho\nu}A^\mathit{el}_{\mu}\partial_{\rho}A^\mathit{em}_{\nu}+\nonumber\\
&+&J^{\mu}A^\mathit{em}_{\mu}+J^{\mu}_{el}A^\mathit{el}_{\mu},\label{effaction}
\end{eqnarray}
where we have added to the Chern--Simons action the external sources:
$J^{\mu}$ is the total charge density current that naturally couples to the electromagnetic field, and $J^{\mu}_{el}$ is a classically conserved current associated to the elastic field $A^{el}_{\mu}$.
Notice that the standard Chern--Simons term bilinear in $A^{em}$ or $A^{el}$ vanish for the total action due to the opposite value of $C_K$ at the two valleys. Only the mixed term survives.

From Eq.~(\ref{effaction}) we can immediately read out the total
charge current density ($\mathcal{S}_\mathit{eff}=\int
d^{3}\mathcal{L}_\mathit{eff}$):
\begin{equation}
\langle J^{i}\rangle =\frac{\delta\mathcal{S}_{eff}}{\delta A^{em}_i}= 2 e\hat{\beta} C_{K}\varepsilon^{ij}\dot{A}^{el}_{j}\equiv 2 e\hat{\beta}\frac{m}{|m|} \varepsilon^{ij}E^{el}_{j},\label{chargeresponse2}
\end{equation}
where we have assumed for simplicity that $A^{el}_0=0$ and have replaced $\varepsilon^{ij0}=\varepsilon^{ij}$.

Equation~(\ref{chargeresponse2}) is the main result of this work.
A non vanishing net charge current density can be obtained as a
response to a time dependent elastic deformation of the gapped
graphene sample. Notice that this equation is consistent with time
reversal symmetry because, as we emphasized previously, the
synthetic electric field is odd under time reversal.~\footnote{We
thank A. G. Grushin for extensive discussions around this point.}

This result is the counterpart of the quantum valley Hall effect.
In Fig.~\ref{fig1} we represent schematically a comparison between
the Hall effect, the valley Hall effect, and the effect proposed
in the text.  The upper part (a) shows the original implementation
of the experiment first proposed by Hall.~\cite{Hall} In the
presence of a magnetic field perpendicular to the sample and a
voltage difference along the x direction the charge carriers are
deflected to one of the edges of the sample. The charge
accumulation in one side gives a voltage difference between the
two sample's edges in the y direction. In the valley Hall effect
(middle part of the figure) there is no magnetic field. The
electromotive force acts along the x direction, but now the
electrons crossing the sample find two channels to flow, one at
each sample's edge. The voltage difference between the two edges
is now zero, but the carriers flowing along different edges belong
to different valleys or Fermi points so there is a net valley
imbalance between the edges in the y direction. Finally the charge
effect proposed in this work is shown in the lower part of the
figure. The external probe now is a time dependent elastic
deformation creating a synthetic electric field along the x
direction. Electrons belonging to different valleys react
oppositely to this electromotive elastic force, so the two
available channels belong to the same edge and a net charge
accumulation occurs in one of the two sides of the sample. Hence a
net voltage difference appears between the edges. Because both
channels belong to different Fermi points, no valley imbalance
appears in this situation.

\begin{figure}[h]
\center{
\includegraphics[width=6.0cm]{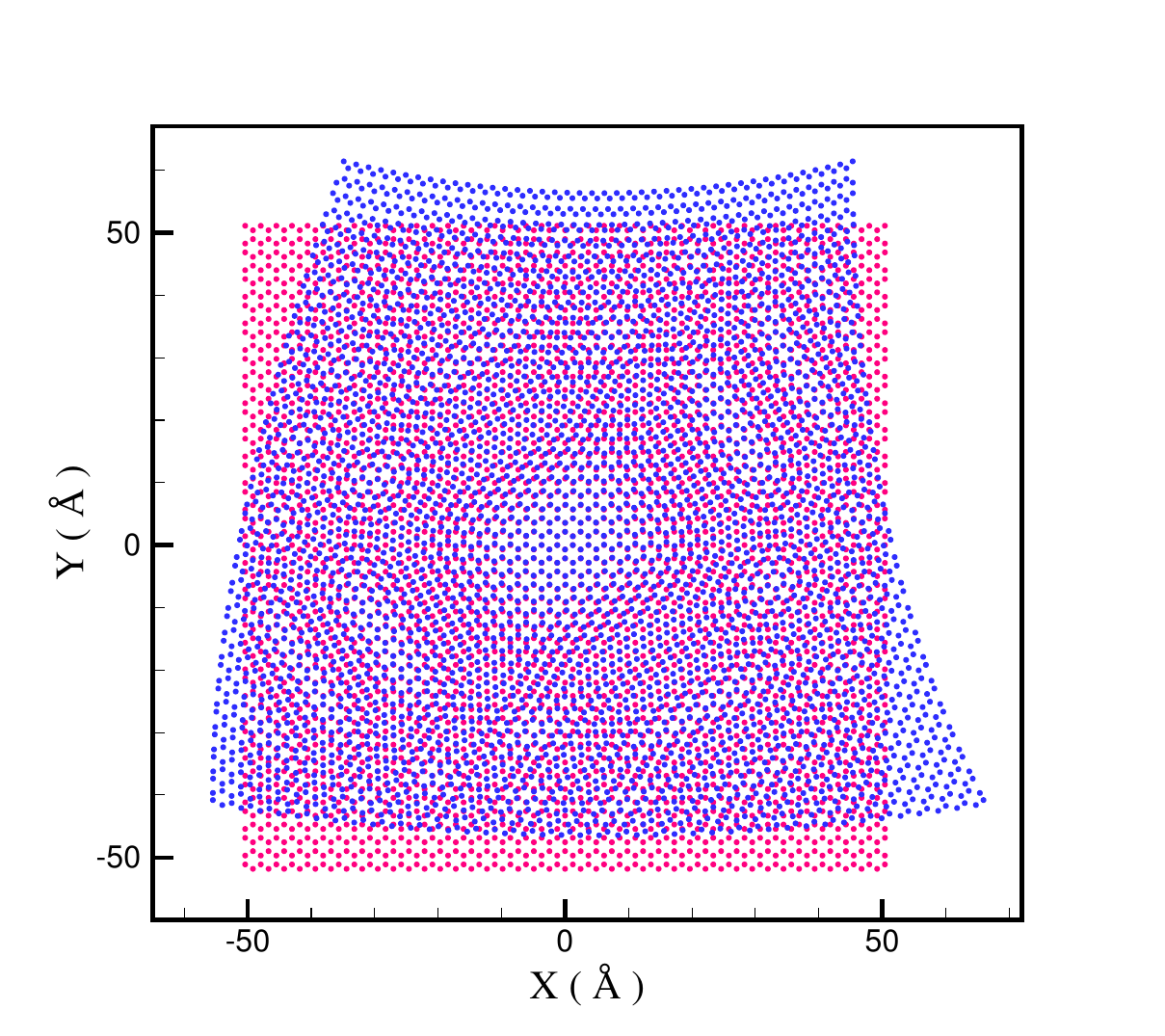}
\includegraphics[width=2.2cm]{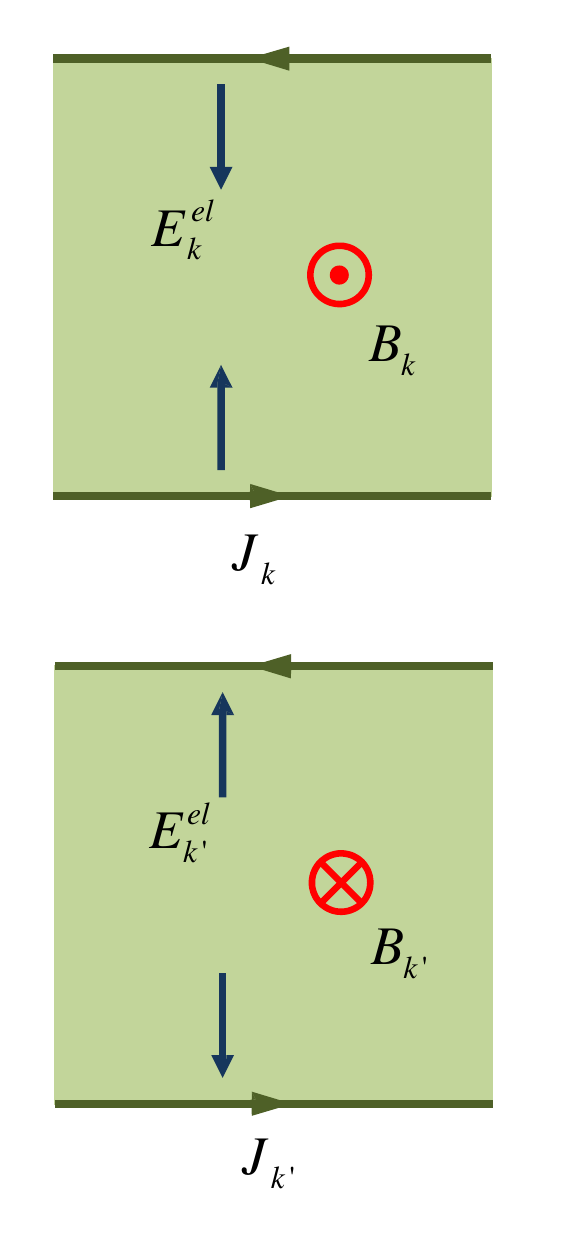}
}
\caption{(Color online) (Left): Proposed strain for a graphene ribbon as described in the main text. (Right): Direction of the charge current for the two Dirac points in different edges.}\label{fig02}
\end{figure}
\section {Suggested experimental realization}

A potential experimental setup to measure the effect described in
this work needs a gapped graphene system with reasonably well
defined zig-zag edges. We also need to induce time dependent
strain but there is no need for high control on this part. One
possibility is to use the proposal of
Ref.~[\onlinecite{Baeetal10,KZetal09}] where graphene is grown on
a thin copper substrate with outstanding flexibility.  The strain
on the graphene sample can be controlled by manipulating the
substrate. The sample can be gapped by chemical doping as done
in~\cite{ENetal09} although the needed gap could also be induced
by the strain field as described in Ref.~[\onlinecite{GKG10}].

The vector field induced by an elastic deformation of the graphene
lattice is given in Eq.~\eqref{gaugefield}. A possible strain
configuration is shown in Fig.~\ref{fig02}. The left hand side
shows the atomic displacements  given by $(u_x,u_y)=(-2
P_B~xy+P_E(t)y^2,P_B(x^2+y^2))$ in  Cartesian coordinates. $P_B$
and $P_E(t)$ are geometric parameters with units of $1$/length and
$P_E(t)$ is  a time-varying periodic function, for instance
$P_E(t)=P_B\cos(\omega t)$. This particular  strain leads to a
uniform magnetic field $\tilde{B}_z=4\phi_0 c\beta P_B/a$ and
pseudo-electric field $\tilde{E}_y=-2\phi_0 c \beta \omega
\sin(\omega t) y /a $. To simplify the analysis the strain has
been chosen such that the induced scalar potential which is
proportional to $u_{xx}+u_{yy}$ vanishes. Neither this condition
nor the uniform pseudomagnetic field are necessary for the
proposed mechanism to work but they provide a neat setting for the
discussion. A value of $P_B= 0.5~(\mu m)^{-1}$ in a sample of size
$0.4~\mu m \times 0.4 \mu m$ gives a pseudo-magnetic field of
$\tilde{B}_{z}\approx 9.0~T$ which is large enough to give rise to
quantized Landau levels. The sample can be tailored in the form of
a ribbon with zigzag termination such that there will be edge
states with a good valley number.~\cite{LG10} The experimental
possibility tailoring proper edges has been demonstrated in
Ref.~[\onlinecite{CMetal09}]. As we discussed in the text, an
external electric field can not induce a current along the
horizontal boundaries, but the pseudo-electric filed given by
$-\partial {\bf A}_{el}(t)/\partial t$ creates easily a voltage
along the boundaries as shown  in the right hand side of
Fig.~~\ref{fig02}. Using a reasonable value for the parameters and
$\omega=60~MHz$, the maximum pseudo-electric field reaches
$50~V/m$ and then $J_x$ would be about $4\times10^{-3}~A/m$. This
induced current can be measured by experiments.

Bilayer graphene~\cite{NCetal06} is another possible system -
perhaps better than the monolayer - to observe the proposed
current. Gated graphene bilayer is known to be another realization
of a Quantum Valley Hall insulator, whose low energy theory is
exactly the same mixed Chern Simons theory described in our work.
The system is gapped when a gate voltage is applied between the
two graphene layers  and it supports the same structure of valley
resolved edge states. As it happens to its monolayer counterpart,
having zigzag edge states is essential to develop such edge
states. However, contrary to the case of monolayer graphene, it
has been experimentally reported a subgap conduction in gated
graphene bilayer ribbons.~\cite{LMetal11} This subgap conductance
has been attributed to the presence of edge states in the sample
indicating that the conduction along the edge states survive even
when there is no perfect zigzag crystalline ordering at the edges
\cite{VLetal13}. Quantum manipulation of valleys in bilayer
graphene has been reported recently.~\cite{WLC13}

\section { Conclusions}
We have shown that a non vanishing charge current can be generated
in gapped graphene by applying time--dependent strain. Unlike the
standard Hall effect proportional to $e^2$, its coefficient is
proportional to the product of the electric charge times an
elastic constant characteristic of graphene, $\hat{\beta}$. The
proposed mechanism is a consequence of the mixed responses that
can be obtained in non--trivial topological Dirac systems when
several vector fields are coupled. The proposed effect can be
measured in actual graphene devices or in alternative systems such
as artificial graphene~\cite{GMetal12} or optical
lattices.~\cite{LGetal09}

\section {Aknowledgements} We wish to thank A. Concha, M. Barkeshli and A. G. Grushin for useful discussions. This research was supported in part by the Spanish MECD grants FIS2011-23713 and PIB2010BZ-00512.

\newcommand{\npb}{Nucl. Phys. B}\newcommand{\adv}{Adv.
  Phys.}\newcommand{\epl}{Europhys. Lett.}

\end{document}